\documentclass[prc,preprint,showpacs,tightenlines,%
                floatfix,amsfonts,nofootinbib]{revtex4}
\usepackage{latexsym}
\usepackage{graphicx}
\usepackage{amssymb}
\usepackage{amsmath}
\usepackage{amscd}

\newcommand{\ugamma}[1]{\gamma^{#1}}

\newcommand{\usigma}[1]{\sigma^{#1}}

\newcommand{\ket}[1]{\vert#1\rangle}
\newcommand{\bra}[1]{\langle#1\vert}

\newcommand{\psibar}[1]{\overline{#1}}
\newcommand{\half}{\ensuremath{\frac{1}{2}}}

\newcommand{\modular}[2]{\vert \vec{#1}_{#2}\vert}

\def\Tr{\mathop{\rm Tr}\nolimits}
\def\Mstar{M^{\ast}}
\def\mstar{m^{\ast}}

\begin{document}

\title{Coherent Neutrinoproduction of Photons and Pions in a Chiral 
            Effective Field Theory for Nuclei}

\author{Xilin Zhang}\email{xilzhang@indiana.edu}
\author{Brian D. Serot} \thanks{Deceased.}
\affiliation{Department of Physics and Center for Exploration of
             Energy and Matter\\
             Indiana University, Bloomington, IN\ \ 47405}
                         
%
\author{\null}
\noaffiliation

%
\date{\today\\[20pt]}

\begin{abstract}
\begin{description}
\item[Background:] The neutrinoproduction of photons and pions from nucleons and nuclei is relevant to the background analysis in neutrino-oscillation experiments [for example, MiniBooNE; A. A. Aquilar-Arevalo \textit{et al.} (MiniBooNE Collaboration), Phys.\ Rev.\ Lett.\ {\bf 100}, 032301 (2008)]. 
The production from nucleons and incoherent production with $E_{\nu} \leqslant 0.5 \ \mathrm{GeV}$ have been studied in [B. D. Serot and X. Zhang, Phys.\ Rev.\ C {\bf 86}, (2012) 015501; and X. Zhang and B. D. Serot, Phys.\ Rev.\ C {\bf 86}, (2012) 035502].
\item[Purpose:] Study coherent productions with $E_{\nu} \leqslant 0.5 \ \mathrm{GeV}$. Also address the contributions of two contact terms in Neutral Current (NC) photon production that are partially related to the proposed anomalous $\omega (\rho)$, $Z$ boson and photon interactions.
\item[Methods:] We work in the framework of a Lorentz-covariant effective field theory (EFT), which contains nucleons, pions, the Delta (1232) ($\Delta$s), isoscalar scalar ($\sigma$) and vector ($\omega$) fields, and isovector vector ($\rho$) fields, and incorporates a nonlinear realization of (approximate) $\mathrm{SU(2)}_{\mathrm{L}} \otimes \mathrm{SU(2)}_{\mathrm{R}}$ chiral symmetry. A revised version of the so-called ``optimal approximation'' is applied, where one-nucleon interaction amplitude is factorized out and the medium-modifications and pion wave function distortion are included. The calculation is tested against the coherent pion photoproduction data. 
\item[Results:] The computation shows an agreement with the pion photoproduction data, although precisely determining the $\Delta$ modification is entangled with one mentioned contact term. The uncertainty in the $\Delta$ modification leads to uncertainties in both pion and photon neutrinoproductions. In addition, the contact term plays a significant role in NC photon production. 
\item[Conclusions:] First, the contact term increases NC photon production by  $\sim 10\%$ assuming a reasonable range of the contact coupling, which however seems not significant enough to explain the MiniBooNE excess. A high energy  computation is needed to gain a firm conclusion and will be presented elsewhere. Second, the behavior of coherent neutrinoproductions computed here is significantly different from the expectation at high energy by ignoring the vector current contribution.  
\end{description}
\end{abstract}

\smallskip

\maketitle

\section{Introduction} \label{sec:intro}

This is the third of a series of studies about neutrinoproduction of photons 
and pions with neutrino energy $E_{\nu}\leqslant 0.5$ GeV where the 
$\Delta$ excitation is important \cite{bookchapter,1stpaper,2ndpaper}. 
The focus of this paper is the coherent production. As we know, in the neutrino-oscillation experiments, for example MiniBooNE \cite{MiniBN2007,MiniBN2009, MiniBN2010}, the photon and pion 
neutrinoproduction from nuclei and nucleons are potential backgrounds.  
It is still a question whether NC photon production might explain the excess events seen at low \emph{reconstructed} neutrino energies in the MiniBooNE experiment, which the MicroBooNE experiment plans to answer \cite{MicroBN2011}. In Refs.~\cite{Harvey07, Harvey08, Hill10, Gershtein81}, 
the authors argued that the anomalous interaction terms involving $\omega (\rho)$, $Z$ boson, and the photon may increase neutral current (NC) photon production. So the cross-section calculation for these processes becomes necessary.

In Ref.~\cite{bookchapter}, we introduce the $\Delta$ resonance as a manifest degrees of freedom in a Lorentz-covariant effective field
theory (EFT) with a nonlinear realization of the $\mathrm{SU(2)}_{\mathrm{L}} \otimes \mathrm{SU(2)}_{\mathrm{R}}$ chiral symmetry \footnote{The EFT was originally motivated by the nuclear many-body problem \cite{SW86,SW97,FST97,FSp00,FSL00,EvRev00,LNP641,EMQHD07}, and is often called \emph{quantum hadrodynamics} or QHD.}. 
In Ref.~\cite{1stpaper}, we study both neutrinoproductions from free nucleons
and calibrate our theory. Because of various symmetries that are built in, the conservation of vector current (CVC), and the partial conservation of axial current (PCAC) are satisfied automatically, which is crucial for photon production calculation. In Ref.~\cite{2ndpaper}, we work on the incoherent productions from the nucleus. The previous studies show that the contributions due to the two terms mentioned above \cite{Hill10} are tiny in the NC photon production from both free nucleons and nuclei. This paper is devoted to the study of coherent productions from ${}^{12}C$, which is the major target nucleus in the MiniBooNE's detector, and also to addressing the significance of the two mentioned terms. 

Here we apply the so-called ``optimal'' approximation, in 
which one-nucleon interaction amplitude can be factorized out from the full 
nuclear matrix element leading to great simplification of the calculation. Meanwhile both the CVC and the PCAC are preserved. The nuclear ground state is calculated by using the mean-field approximation (see Ref.~\cite{FST97} for the details). The optimal approximation was first illustrated 
generally for projectile-nucleus scattering in Refs.~\cite{Gurvitz79jan, Gurvitz79oct}. It has been applied quite successfully to nucleon-nucleus 
scattering in a relativistic framework \cite{McNeil83prl, Shepard83, McNeil83prc, Clark83}. Moreover, a similar approximation has been applied in pion-nucleus elastic scattering \cite{Landau73, Mach83, Gmitro85} and coherent pion photo- and electro-production \cite{Tiator80, Chumbalor87, Eramzhyan90, 
Carrasco93, Peters98, Drechsel99, AbuRaddad99}\footnote{Ref.~\cite{AbuRaddad99}
pointed out that using different one-nucleon interaction amplitudes 
that are equivalent on shell can lead to quite different 
results. Here such ambiguity does not exist because we have a unique 
free interaction amplitude. This will be addressed later.}. 
It was realized that the medium-modification of the one-nucleon 
interaction amplitude plays a key role in the coherent production, which is also included in our revised approximation. In the QHD EFT model, baryons interact with each other via. exchanging the mesons (in the space-like region). Because these bosonic fields develop finite expectation values in the medium, the real part of the baryon self-energy is modified on the mean-field level \cite{SW86}. Meanwhile the change of the $\Delta$ width has been studied 
in the nonrelativistic framework both phenomenologically 
\cite{Hirata79, Horikawa80} and theoretically \cite{Oset87}, 
but it is not completed in the relativistic framework 
\cite{wehrgerger89, wehrgerger90, wehrgerger92, wehrgerger93, 2ndpaper}. 
Here we continue our simple treatment proposed in Ref.~\cite{2ndpaper}. Another important factor related to the $\Delta$ 
is the distorted pion wave function in the pion 
production, which is included here by using the Eikonal approximation. Such an  effect can be ignored for photon production. 
Comparing the two may be used to disentangle the medium-modification and the pion wave function distortion.

To benchmark the approximation scheme, we calculate various differential 
cross sections for pion photoproduction. We are able to get an agreement with existing data \cite{Schmitz96, Gothe95, Arends83}. The approximation is then  applied to study the photon and pion neutrinoproduction. Unfortunately, existing neutrino experiments, for example Refs.~\cite{k2k05prl, miniboone08plb}, do not put a strong constraint on pion productions with $E_{\nu}\leqslant 0.5$ GeV, since most of them have only spectrum-averaged measurements, and the mean neutrino energy is around $1$ GeV. 
On the theoretical side, there are other microscopic calculations on pion productions \cite{Singh06, Alvarez-Ruso07CC, Alvarez-Ruso07CCErratum, Alvarez-Ruso07NC, Alvarez-Ruso07NCErratum, Amaro09, Hernandez09, Leitner09, Nakamura10, Gershtein1980,Komachenko1987}. In most of them, the optimal approximation is in one way or another applied. The $\Delta$ dynamics is 
taken into account by using the nonrelativistic models. The final pion wave function is calculated either in the Eikonal approximation or by solving the Schroedinger equation with the pion optical potential. The key difference between our work and others is that we work in a Lorentz-covariant EFT, which has been applied successfully to nuclear many-body problems  and also has been calibrated for neutrinoproductions from free nucleons. The medium-modification of baryons can be calculated on the mean-field level. We can address the power counting of different diagrams in this EFT, although the theory can only be used at the low energy region ($E_{\nu} \leqslant 0.5 \mathrm{GeV}$). More importantly, coherent NC photon production has rarely been discussed in the microscopic approach. In addition, there exists a macroscopic approach, which treats the nucleus as a whole and makes use of the forward scattering behavior of coherent pion production in the high energy scattering. 
In the forward scattering kinematics, PCAC leads to a 
relation between the pion neutrinoproduction and pion-nucleus elastic scattering. 
This is initiated in Refs.~\cite{Adler64, Rein83}, used in the NUANCE event generator \cite{nuance,MBCCQE}, 
and revised recently in Refs.~\cite{Paschos06, Rein07, Berger09}. This approach has also been applied to compute coherent photon neutrinoproduction at the $2$ GeV region and beyond \cite{Rein81}. The NUANCE output on coherent pion production shown throughout this paper, with which we will compare our results, is obtained from the NUANCE v3 event generator with the calibration applied from the experimental data \cite{nuance,miniboone08plb,Sam12} \footnote{The experimental analysis indicates that to be consistent with the coherent NC pion production data, a 35\% reduction needs to be applied to the orginal NUANCE output \cite{miniboone08plb,Sam12}.}. 

The paper is organized as follows. In Sec.~\ref{sec:appscheme}, 
the approximation scheme is discussed, and the difference 
between our approach and others is emphasized. 
Sec.~\ref{sec:results} presents our results. 
At first, differential cross sections for pion photoproduction are 
compared to the data. The effect of the two contact terms are 
discussed. Then, we show the results for pion neutrinoproductions. 
Since there is the uncertainty in our model, results of 
using different parameters are compared. 
Finally, we focus on NC photon production and discuss the relevance of our results to the MiniBooNE low reconstructed energy excess events. The two contact terms are again discussed in this context.

\section{Approximation scheme} \label{sec:appscheme}

\subsection{Kinematics}

The formulas needed for computations are shown here. 
For the $\pi^{0}$ photoproduction,
\begin{eqnarray}
\sigma &=&\int \frac{1}{(8\pi)^{2}} \frac{\modular{k}{\pi}}{E_{\gamma}} \frac{\psibar{\left| M \right|^{2}}}{m_{A}^{2}} d\Omega_{\pi} \ . 
\end{eqnarray}
Here $q^{\mu}$ and $k_{\pi}$ are the momentum of the incoming photon and outgoing pion, and $q^{0}=E_{\gamma}$ is the photon energy in the laboratory frame. Because the nucleus $A$ remains in the ground state and is heavy enough 
to ignore its recoil, we can have $k_{\pi}^{0}=q^{0}$. The $1/m_{A}$ term is used to properly normalize the quantum state.  
The definition of transition probability is 
\begin{eqnarray}
\frac{\psibar{\left| M \right|^{2}}}{m_{A}^{2}} &=& \frac{1}{m_{A}^{2}} \half \sum_{\lambda_{i}} e^{2} \left| \epsilon_{\mu}(\lambda_{i} \vec{q}) \bra{A, \pi (\vec{k}_{\pi})} J_{had}^{\mu}\ket{A} \right|^{2} \ ,
\end{eqnarray}
where $J_{had}^{\mu}$ is the Electromagnetic current involved in this process, and $\lambda_{i}$ is the photon polarization. 

For the neutrinoproduction,
\begin{eqnarray}
\sigma&=&\int \frac{1}{(4\pi)^{5}} \frac{\modular{k}{\pi}\modular{p}{lf}}{E_{\nu}} \frac{\psibar{\left| M \right|^{2}}}{m_{A}^{2}} dE_{lf} d\Omega_{lf}d\Omega_{\pi} \ . \label{eqn:xsectionpion}
\end{eqnarray}
We define $p_{li}$ and $p_{lf}$ as the momenta of incoming and outgoing leptons. 
$q\equiv p_{li}-p_{lf}$, $p_{li}^{0}=E_{\nu}$ (the incoming lepton energy in 
the laboratory frame). Here we also have $k_{\pi}^{0}=q^{0}$. The nuclear matrix element is 
\begin{eqnarray}
\frac{\psibar{\left| M \right|^{2}}}{m_{A}^{2}}&=&  \frac{1}{m_{A}^{2}}\sum_{s_{lf}} 
(4\sqrt{2}V_{ud}G_{F})^{2} \left| l_{\mu}(\vec{p}_{li}, \vec{p}_{lf}) \bra{A, \pi (\vec{k}_{\pi})} 
J_{had}^{\mu}\ket{A} \right|^{2} \ . \label{eqn:currentpion}
\end{eqnarray}
Here $G_{F}$ is the Fermi constant; $V_{ud}$ is the $u$ and $d$ quark mixing 
in the charged current (CC), and is $1$ in the NC; and $l_{\mu}(\vec{p}_{li}, \vec{p}_{lf})$ is 
the corresponding lepton current.
Moreover, 
$J_{had}^{\mu} = J_{EM}^{\mu}=\half J_{B}^{\mu}+V^{0\mu}$ for the 
photoproduction, $J_{had}^{i\mu}= \half(V^{i\mu}+A^{i\mu}), i=\pm 1 $ 
for the CC, and 
$J_{had}^{\mu}= \half(V^{0\mu}+A^{0\mu})-\sin^{2}\theta_{w}J_{EM}^{\mu}$ 
for the NC ($\theta_{w}$ is the weak mixing angle). 

For NC photon production, the zero mass of photon should be taken into 
account in Eq.~(\ref{eqn:xsectionpion}), 
and $\bra{A, \pi} J_{had}^{\mu}\ket{A}$ in Eq.~(\ref{eqn:currentpion})
needs to be changed to $\bra{A, \gamma} J_{had}^{\mu}\ket{A}$.

\subsection{The optimal approximation} \label{subsec:approximationscheme}

The current matrix element can be written as 
\begin{eqnarray}
&&{} \frac{1}{m_{A}} \bra{A, \pi (\vec{k}_{\pi})} J_{had}^{\mu}\ket{A}  \notag \\[5pt]
&\approx& 
\begin{cases} 
\int_{A} d \vec{r}e^{i(\vec{q}-\vec{k}_{\pi})\cdot\vec{r}} \langle J_{had}^{\mu}(\vec{q},\vec{k}_{\pi}, \vec{r}) \rangle & \mathrm{PW}  , \\[8pt] 
\int_{A} d \vec{r}e^{i(\vec{q}-\vec{k}_{\pi})\cdot\vec{r}} e^{-i\int^{\infty}_{z} \frac{\Pi(\rho,l)} {2\modular{k}{\pi}}dl} \langle J_{had}^{\mu}(\vec{q},\vec{k}_{\pi}, \vec{r}) \rangle
& \mathrm{DW} . 
\end{cases} \label{eqn:onebodymatixelement}
\end{eqnarray}
Only the one-body current contributions are included coherently. 
We apply the optimal
approximation to simplify the calculation \cite{Tiator80, Chumbalor87, Eramzhyan90, 
Carrasco93, Peters98, Drechsel99, AbuRaddad99}:
\begin{align}
\langle J_{had}^{\mu}(\vec{q},\vec{k}_{\pi}, \vec{r}) \rangle \approx & \ \rho_{n}(\vec{r}) \half \sum_{s_{z}}\frac{1}{p_{ni}^{\ast 0}}\bra{n,s_{z}, \frac{\vec{q}-\vec{k}_{\pi}}{2}} J_{had}^{\mu}(\vec{q},\vec{k}_{\pi})\ket{n, s_{z}, \frac{\vec{k}_{\pi}-\vec{q}}{2}} \notag \\
&+\rho_{p}(\vec{r})\half \sum_{s_{z}}\frac{1}{p_{ni}^{\ast 0 }} \bra{p, s_{z}, \frac{\vec{q}-\vec{k}_{\pi}}{2}} J_{had}^{\mu}(\vec{q},\vec{k}_{\pi})\ket{p, s_{z},\frac{\vec{k}_{\pi}-\vec{q}}{2}} \ . \label{eqn:factorization}
\end{align} 
Refs.~\cite{Gurvitz79jan, Gurvitz79oct} argued that in the center mass frame 
of the projectile and the \emph{nucleus}, the nuclear matrix element can be expressed as the product of a proper density and the free one-nucleon interaction
amplitude calculated in the Breit frame of the projectile and the \emph{nucleon}. 
Ignoring the recoil of the nucleus leads to Eq.~(\ref{eqn:factorization}). 
For NC photon production, 
we can use Eqs.~(\ref{eqn:onebodymatixelement}) and 
(\ref{eqn:factorization}) with a proper current inserted. 
 
\begin{figure}
\centering
\includegraphics[scale=0.6]{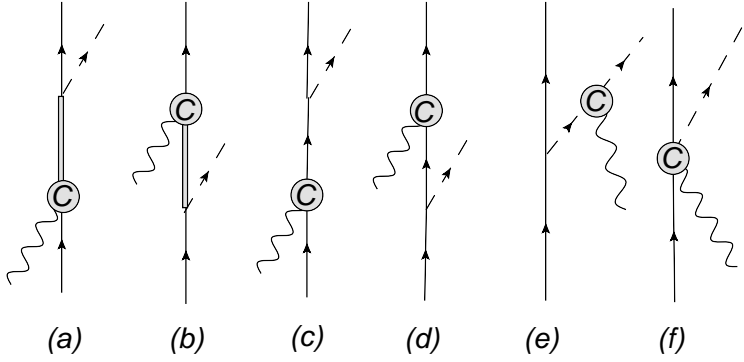}
\caption{The Feynman diagrams for pion production. Here {\bf C} stands for various types of currents including vector, axial-vector, and baryon currents. Some diagrams may be zero for some specific type of current. See Ref.~\cite{1stpaper} for the details.}
\label{fig:feynmanpionproduction}
\end{figure}

\begin{figure}[!h]
\begin{center}
\includegraphics[scale=0.7, angle=-90]{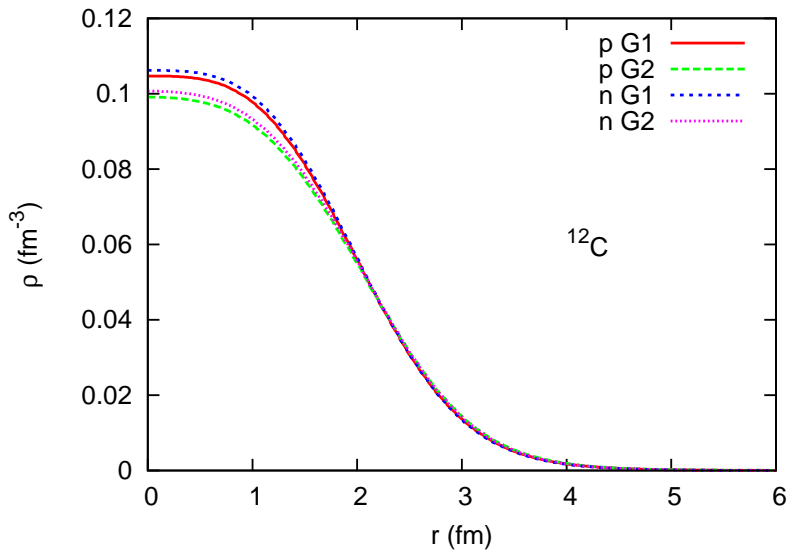}
\caption{(Color online). The proton and neutron densities in ${}^{12}C$ as calculated in the mean-field approximation by using G1 and G2 parameter sets \cite{FST97}.}
\label{fig:c12density}
\end{center}
\end{figure}

\begin{figure}[!h]
\begin{center}
\includegraphics[scale=0.7, angle=-90]{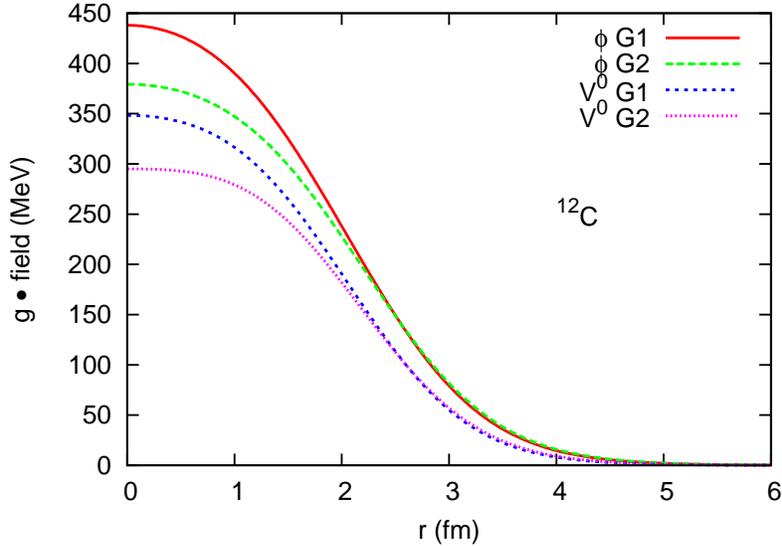}
\caption{(Color online). The field expectation values in ${}^{12}C$, $\langle g_{s} \phi \rangle$ and $\langle g_{v} V^{0}\rangle$, as calculated in the mean-field approximation by using G1 and G2 parameter sets \cite{FST97}.}
\label{fig:c12fields}
\end{center}
\end{figure}

The calculation of the one-body current matrix element for both pion and photon production in Eq.~(\ref{eqn:factorization}) has been discussed in Refs.~\cite{1stpaper,2ndpaper}. There are two basic types of Feynman diagrams contributing here, as shown in Fig.~\ref{fig:feynmanpionproduction}: diagrams with the $\Delta$ [(a) and (b)] and all the rest called nonresonant diagrams here. The diagrams for the photon production can be viewed as those in Fig.~\ref{fig:feynmanpionproduction} with the final pion line changed to the photon line. 
The medium modification on the one-nucleon interaction amplitude, as introduced in Ref.~\cite{2ndpaper}, is based on the mean-field approximation. The effective mass is introduced for the baryon to include the modification on the real part of its self-energy:
\begin{eqnarray}
\Mstar &\equiv& M-g_{s} \langle \phi \rangle \label{eqn:bgcoupling3} \ ,\\[5pt]
\mstar &\equiv& m-h_{s} \langle \phi \rangle \label{eqn:bgcoupling4}\ , \\[5pt]
p_{n}^{0}&\equiv& p_{n}^{\ast 0}+ g_{v}\langle V^{0}\rangle=\sqrt{M^{\ast2}+\vec{p}_{n}^{2}} + g_{v}\langle V^{0}\rangle \ , \label{eqn:bgcoupling1} \\[5pt]
p_{\Delta}^{0}&\equiv& p_{\Delta}^{\ast 0}+ h_{v}\langle V^{0}\rangle=\sqrt{m^{\ast 2}+\vec{p}_{\Delta}^{2}}+h_{v}\langle V^{0}\rangle \ . \label{eqn:bgcoupling2} 
\end{eqnarray}
Here $g_{s,v}$ ($h_{s,v}$) are the couplings between the scalar and vector mesons and the nucleon (the $\Delta$). 
Figs.~\ref{fig:c12density} and \ref{fig:c12fields} show the 
calculated $g_{s} \langle \phi \rangle$ and $g_{v}\langle V^{0}\rangle$ in ${}^{12}C$ (we approximate it as a spherical nucleus). ``G1'' and ``G2'' label two parameter sets about $g_{s}$,  $g_{v}$, and others \cite{FST97}. In this paper, we use ``G1'' as in Ref.~\cite{2ndpaper}. 
For the $\Delta$ width, we follow Refs.~\cite{2ndpaper, Hirata79, Horikawa80}. 
Above the pion threshold, 
\begin{eqnarray}
\Gamma_{\Delta}&=&\Gamma_{\pi} + \Gamma_{\mathrm{sp}}  \ , \notag \\
\Gamma_{\mathrm{sp}}&\approx&V_{0}\times \frac{\rho(r)}{\rho(0)} \ , \notag  
\end{eqnarray}
where $\Gamma_{\pi}$ is the $\Delta$ pion-decay width in the nucleus, and can be found in Refs.~\cite{wehrgerger92, Oset87, Nieves93}. $\Gamma_{\mathrm{sp}}$ is the width of other channels, which 
has been fitted in Refs.~\cite{Horikawa80, Nakamura10}. 
We set $V_{0} \approx 80 \ \mathrm{MeV}$ 
\cite{2ndpaper, Horikawa80, Nakamura10}. 
$\rho(r)$ is the baryon density at radius $r$. 
Below the pion threshold in the photon production, 
\begin{eqnarray}
\Gamma_{\Delta}\approx\Gamma_{\mathrm{sp}}\approx V_{0}\times \frac{\rho(r)}{\rho(0)} 
\ . \notag
\end{eqnarray}
In the cross channel of the $\Delta$ diagram, we set the width 
to be zero.

In addition, since pions interact strongly with the nucleus, 
it is necessary to treat the final pion wave function in a realistic way. 
As shown in Eq.~(\ref{eqn:onebodymatixelement}), the Eikonal approximation is used to calculate the distorted wave function \cite{Singh06}, which is labeled as DW, while the PW calculation is without such distortion.
For NC photon production, we only apply the PW calculation. 
In Eq.~(\ref{eqn:onebodymatixelement}), $\Pi(\rho(\vec{r}), z)$ is the pion polarization insertion in the nuclear medium with baryon density $\rho(\vec{r})$, as calculated in the local Fermi gas approximation. Following Refs.~\cite{Alvarez-Ruso07CC, Alvarez-Ruso07NC, Singh06}, 
we use the following formula for $\Pi$ in symmetric nuclear matter:
\begin{eqnarray}
\Pi&=&-4\pi \frac{M^{2}}{s} \vec{k}_{\pi}^{2} \frac{\mathcal{P}}{1+4\pi g^{\prime} \mathcal{P}}
 \ , \notag \\
\mathcal{P}&=& -\frac{1}{9\pi} \rho \left(\frac{h_{A}}{f_{\pi}}\right)^{2} \Big[( \ \sqrt{s}-m-\mathrm{Re}\Sigma_{\Delta0}+i \Gamma_{\pi}/2-i \mathrm{Im} \Sigma_{\Delta})^{-1}  \notag \\ 
&&{} +  (-\sqrt{s}-m+2M-\mathrm{Re}\Sigma_{\Delta0})^{-1} \Big ] \ , 
\label{eqn:opticalpot}
\end{eqnarray}
where $g^{\prime}=0.63$, $\Sigma_{\Delta}$ is the $\Delta$ self-energy insertion, and $\Gamma_{\pi}$ is the $\Delta$ pion decay width 
as discussed before. We take the results from Ref.~\cite{Alvarez-Ruso07CC} for the $\Sigma_{\Delta}$ and $\mathrm{Re}\Sigma_{\Delta0}$ (See Refs.~\cite{Salcedo88, Nieves93} for the  details) \footnote{We essentially 
treat Eq.~(\ref{eqn:opticalpot}) as an analytical expression for the pion 
optical potential, but it is not used to deal with the $\Delta$ modification
in the one-nucleon interaction amplitude.}. 

\subsection{The approximation used in Ref.~\cite{AbuRaddad99}} \label{subsec:AbuRaddad99}

\begin{figure}
\includegraphics[scale=0.7, angle=-90]{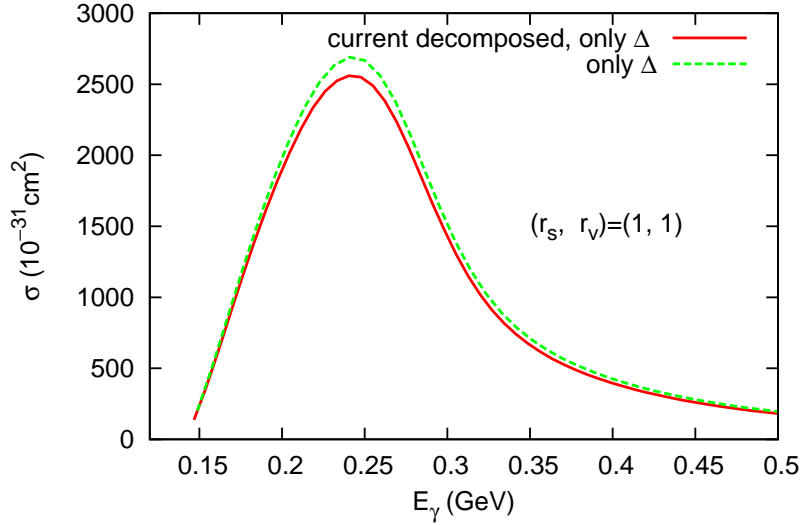}
\caption{(Color online). The photon energy dependence of the total cross section for coherent $\pi^{0}$ photoproduction from ${}^{12}C$. In both calculations, we set $(r_{s}, \ r_{v})=(1, \ 1)$, which controls the $\Delta$ medium modification. The explanation for different calculations can be found in the text.}
\label{fig:compared_with_currentdecomposed}
\end{figure}

It is interesting to compare our calculation with 
that in Ref.~\cite{AbuRaddad99} where the relativistic 
mean-field theory is also used. Instead of using Eq.~(\ref{eqn:factorization}), the authors there project the one-nucleon interaction amplitude to an independent basis, and then convolute the amplitudes of each basis with the corresponding current densities calculated in the relativistic mean-field theory. Take the proton contributions, for instance. First decompose the free proton interaction matrix element: 
\begin{eqnarray}
\bra{p, s_{z}, \frac{\vec{q}-\vec{k}_{\pi}}{2}} J_{had}^{\mu}(\vec{q},\vec{k}_{\pi})\ket{p, 
s_{z},\frac{\vec{k}_{\pi}-\vec{q}}{2}}=\psibar{u}_{f} \left( F_{S}^{\mu}+F_{V \alpha}^{\mu} \ugamma{\alpha}
+F_{T \alpha \beta}^{\mu} \usigma{\alpha\beta} + ...\right) u_{i}   \ , \label{eqn:currentdecomp}
\end{eqnarray} 
and then multiply the amplitude, for example $F_{V\alpha}^{\mu}$, with the proton vector current density $\bra{A} \psibar{\psi_{p}} (x)\ugamma{\alpha} \psi_{p}(x) \ket{A}$. The sum of different terms' contributions in Eq.~(\ref{eqn:currentdecomp}) is the proton contribution to the nuclear matrix element. For the closed shell nucleus, the only relevant amplitudes are $F_{S}, F_{V}$ and $F_{T}$, because the densities associated with other amplitudes in Eq.~(\ref{eqn:currentdecomp}) are zero for a spherical nucleus. We have compared the calculations for the pion production by using Eq.~(\ref{eqn:currentdecomp}) with those by using Eq.~(\ref{eqn:factorization}). Fig.~\ref{fig:compared_with_currentdecomposed} shows the comparison of the total cross section of coherent $\pi^{0}$ photoproduction: the ``current decomposed'' uses  Eq.~(\ref{eqn:currentdecomp}), while the other uses Eq.~(\ref{eqn:factorization}). In the two calculations, we 
include the same medium-modification to the one nucleon matrix element 
(Ref.~\cite{AbuRaddad99} uses the free amplitude) and the same Eikonal 
approximation to calculate final pion wave function. For the $\Delta$ medium modification, we set $(r_{s}, \ r_{v})=(1, \ 1)$ [$r_{s}\equiv h_{s}/g_{s}$, $r_{v}\equiv h_{v}/g_{v}$; see Eqs.~(\ref{eqn:bgcoupling3}) to (\ref{eqn:bgcoupling2})].
Only the $\Delta$ diagrams are considered, because including the others 
needs the extra care of the Electromagnetic current conservation in the 
``current decomposed'' calculation. We see the difference between the two is small. As we have checked, this is also true for the differential cross section and for the cross section of the neutrinoproduction. 
 
In addition, it was pointed out in Ref.~\cite{AbuRaddad99} that there exist other amplitudes that have the same on-shell behavior in nucleon scattering but give quite different results for nucleus scattering through using Eq.~(\ref{eqn:currentdecomp}). There is no such ambiguity in our approach, because we have a unique interaction amplitude derived from the QHD EFT Lagrangian. This shows the importance of having a consistent framework describing both nucleon and nucleus scattering. 

\section{Results} \label{sec:results}
\subsection{Coherent $\pi^{0}$ photoproduction} \label{subsec:photoprod}

\begin{figure}
\includegraphics[scale=0.6,angle=-90]{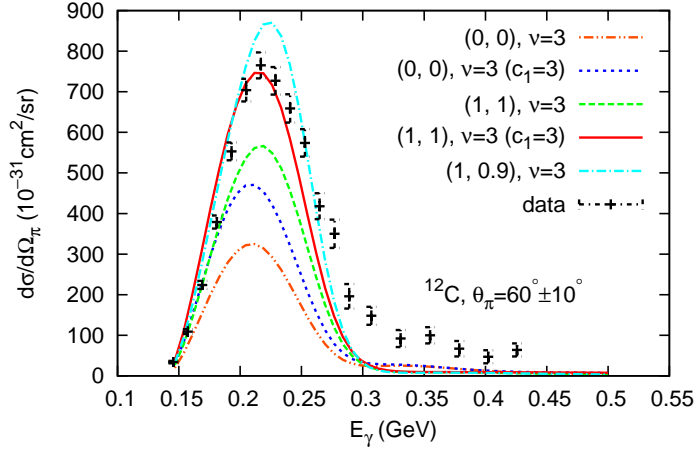}
\caption{(Color online). The photon energy $E_{\gamma}$ dependence of  $d\sigma/d\Omega_{\pi}$ for coherent $\pi^{0}$ photoproduction from ${}^{12}C$. The final pion angle is fixed at $\theta_{\pi}=60^{\circ}\pm10^{\circ}$. The explanation for different calculations can be found in text. The data are from \cite{Schmitz96}.}
\label{fig:photopionprod_theta_60_c12}
\end{figure}

Fig.~\ref{fig:photopionprod_theta_60_c12} shows five different 
calculations for the photon energy dependence of $d\sigma/d\Omega_{\pi}$ of pion photoproduction from ${}^{12}C$ with the pion angle fixed at $\theta_{\pi}=60^{\circ}\pm10^{\circ}$ (relative to the incoming photon direction). All the variables are measured in the laboratory frame of the nucleus.
The data are from Ref.~\cite{Schmitz96}.
These calculations include diagrams up to the $\nu=3$ order ($\nu=2$ terms do not contribute in this production) \cite{1stpaper}. 
Again in the labellings of different curves, $(r_{s}, \ r_{v})$ are defined as $ (h_{s}/g_{s}, \ h_{v}/g_{v})$. Since the two couplings are not precisely known, we simply show results with three different choices: $(0, \ 0)$, $(1, \ 1)$ and $(1, \ 0.9)$ \footnote{It has been shown that  $(r_{s}, \ r_{v})=(0, 0)$ can not explain the $\Delta$ spin-orbit coupling. 
Detailed discussions about the two can be found in \cite{2ndpaper}.}. 
As discussed in Refs.~\cite{bookchapter,1stpaper,2ndpaper}, 
there are two low-energy contact terms involving the photon, 
nucleon, and Z boson (or $\pi$) that contribute at the $\nu=3$ order in NC 
production of the \emph{photon}:
\[\frac{c_{1}}{M^{2}}\psibar{N}\ugamma{\mu} N\Tr \left(\widetilde{a}^{\nu} 
\psibar{F}^{(+)}_{\mu\nu}\right) \ , \ \frac{e_{1}}{M^{2}}\psibar{N}\ugamma{\mu} 
\widetilde{a}^{\nu}N \psibar{f}_{s\mu\nu} \ .\] 
Here $\psibar{F}^{(+)}_{\mu\nu}$ and $\psibar{f}_{s\mu\nu}$ are related 
to the photon field, and $\widetilde{a}^{\nu}$ is related to both Z boson and pion fields. Interestingly in Ref.~\cite{Peters98}, it is shown that $c_{1}$ term plays a significant role in coherent pion photoproduction. 
Refs.~\cite{Harvey07, Harvey08,Hill10} point out the anomalous interactions 
of $\omega$ and $\rho^{0}$ mesons can induce such contact terms at low energy 
with $c_{1}=1.5$ and $e_{1}=0.8$
\footnote{In Ref.~\cite{Peters98}, the authors use the $\omega$'s pion-decay vertex to generate the $c_{1}$ coupling ($c_{1}$ is also around $1.5$). }. 
However, as argued in Ref.~\cite{1stpaper}, $c_{1}$ can also be induced by the 
off-shell interactions involving the $\Delta$, which leaves its value unfixed.
In our calculations shown in Fig.~\ref{fig:photopionprod_theta_60_c12},
we use $c_{1}=1.5$, except for those labeled with $``(c_{1}=3)$'' where we double $c_{1}$. (Since the $e_{1}$ term's contribution 
vanishes for an isospin $0$ target, we focus on $c_{1}$ in the following.)

We can see that the first two calculations with ``$(0 \ , 0)$''  
fail to give the right predictions around the peak. 
``$(1, \, 1), \nu=3 \ (c_{1}=3)$'' and ``$(1, \, 0.9), \ \nu=3$'' give the best predictions. In Ref.~\cite{Peters98}, it is also noticed that fixing the real part of the $\Delta$ self-energy is correlated with $c_{1}$. However when the photon energy is above 0.3 GeV, all the calculations underestimate the cross section. The shapes of different curves are controlled by the nuclear form factor, e.g. the Fourier-transformation of nuclear densities [see Eqs.~(\ref{eqn:onebodymatixelement}) and (\ref{eqn:factorization})], and hence they are similar. So we expect the underestimation to be generic for all the one-body-current calculations. To resolve this issue, two-body currents may need to be considered. In addition, around the peak, the $(0, \  0)$ result is smaller than the $(1, \  1)$ and the $(1, \  1)$ smaller than the $(1, \  0.9)$  ($c_{1}=1.5$ in all the three). The same pattern has been found in the incoherent productions \cite{2ndpaper}. It was argued that among the three, the $(0, \  0)$ requires the most energy to excite the $\Delta$, while the $(1, \  0.9)$ requires the least. In the coherent production, the nuclear form factor makes them even more sensitive to $r_{s}$ and $r_{v}$. This will also be seen for the total cross section of pion neutrinoproduction.

\begin{figure}
\includegraphics[scale=0.6, angle=-90]{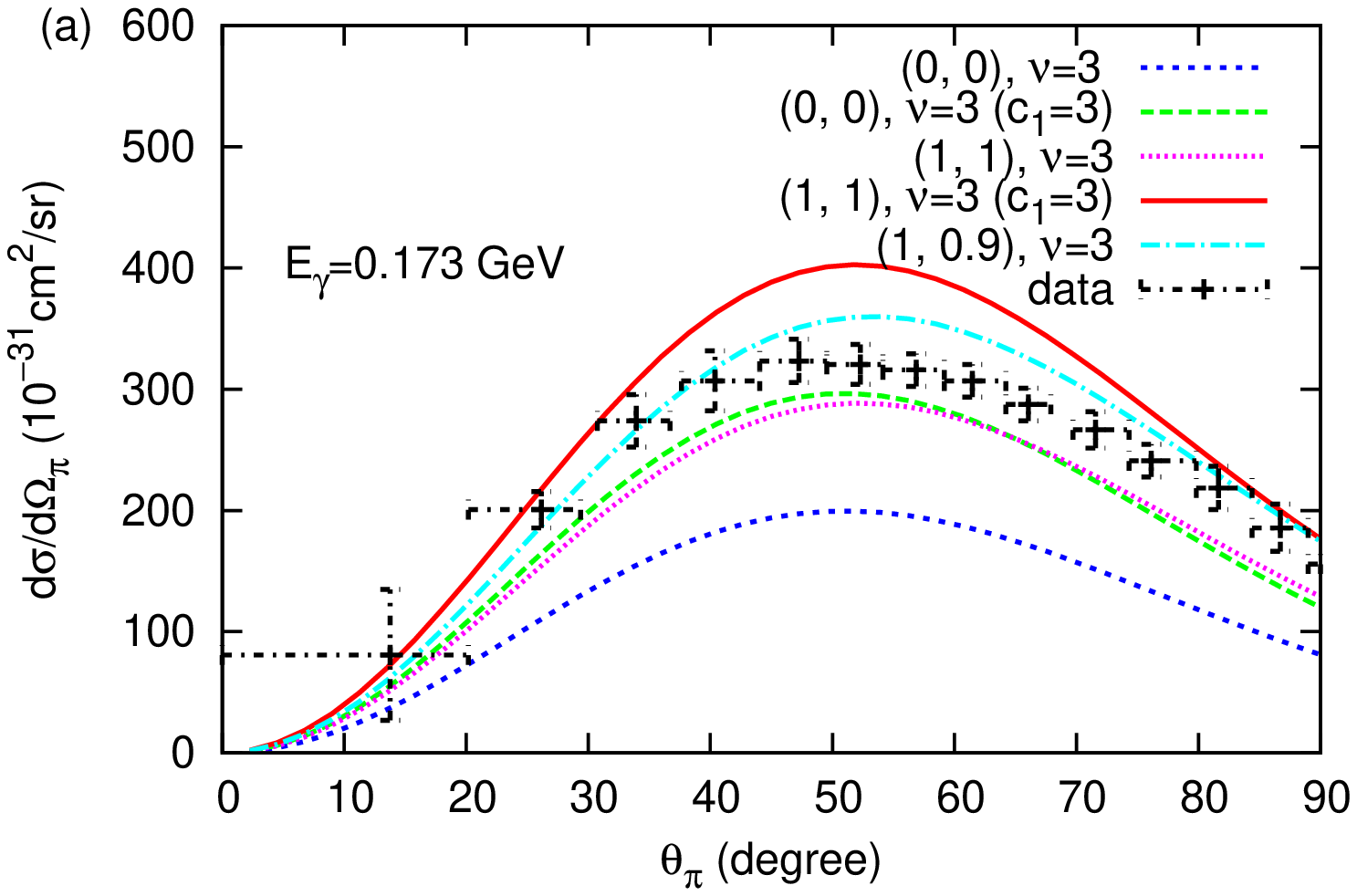}
\includegraphics[scale=0.62, angle=-90]{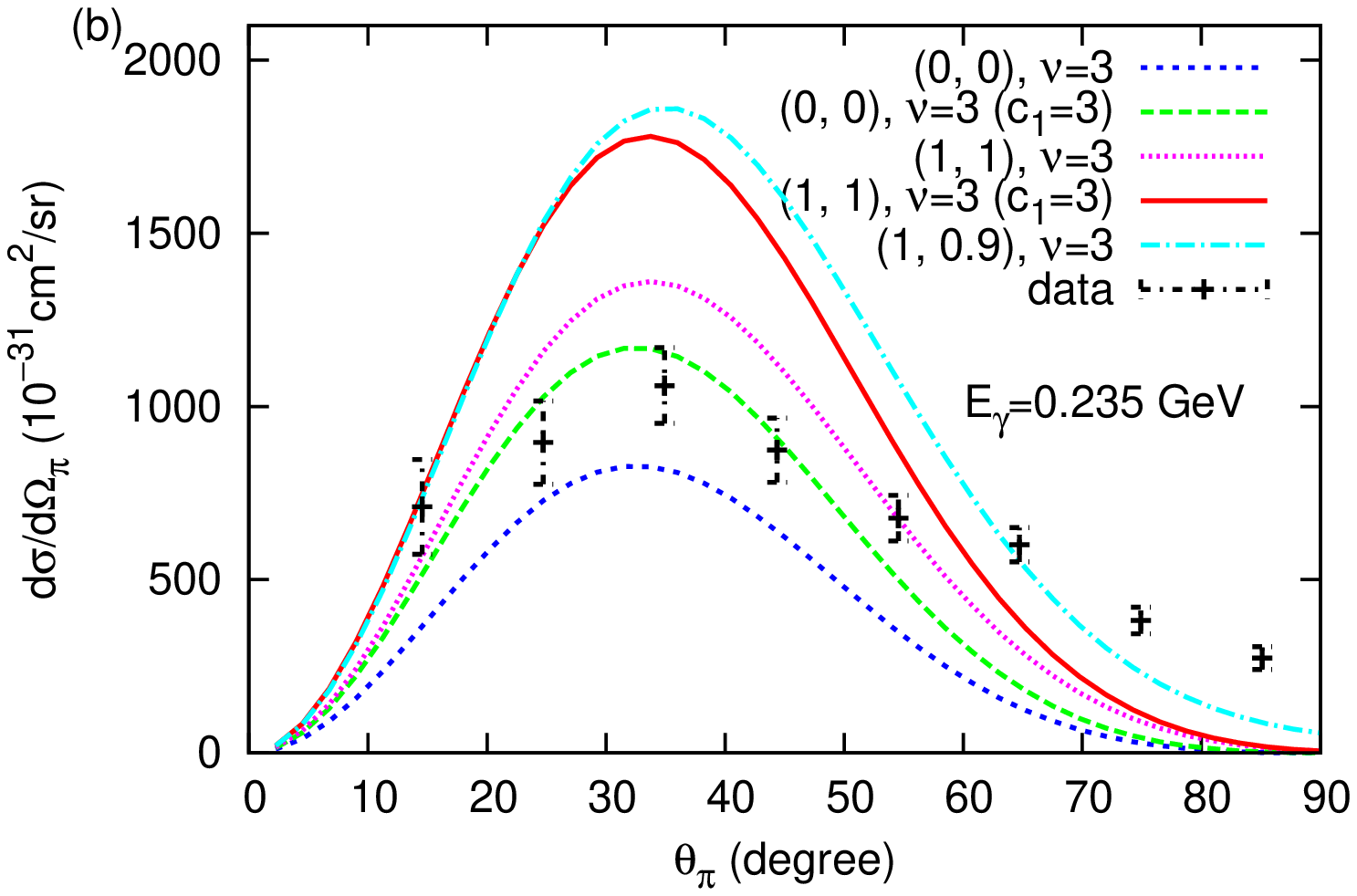}
\includegraphics[scale=0.62, angle=-90]{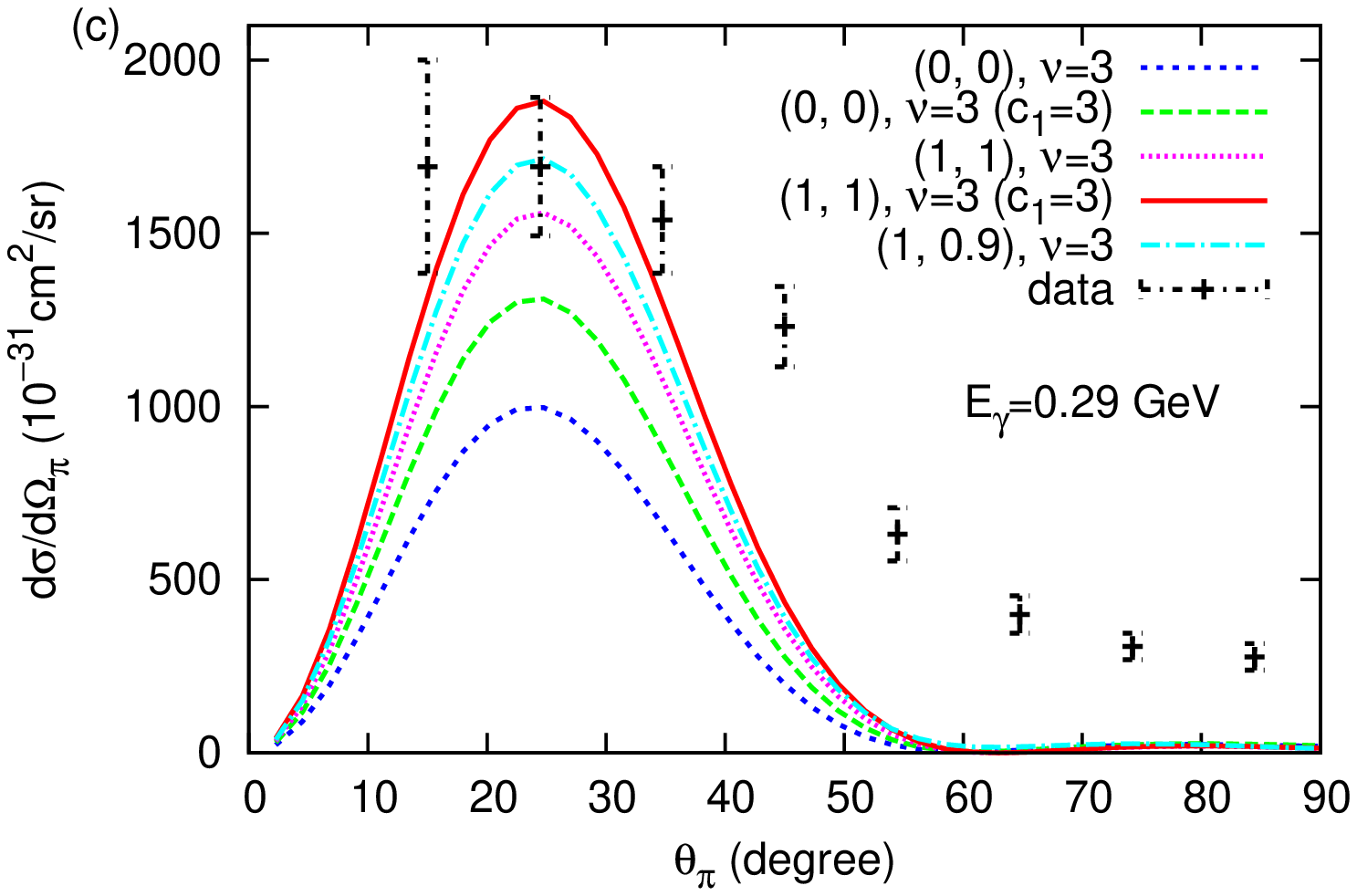}
\caption{(Color online). The angle $\theta_{\pi}$ dependence of $d\sigma/d\Omega_{\pi}$ of coherent $\pi^{0}$ photoproduction from ${}^{12}C$ with the photon energy fixed at $E_{\gamma}=0.173$, $0.235$, and $0.29$ GeV. The explanation for different calculations can be found in the text. The data are from Refs.~\cite{Gothe95, Arends83}.} \label{fig:photopionprod_Ei_c12}
\end{figure}

In Fig.~\ref{fig:photopionprod_Ei_c12}, we show our calculations 
for the scattering angle $\theta_{\pi}$ dependence of 
$d\sigma/d\Omega_{\pi}$ with the photon energy fixed at  
$E_{\gamma}=0.173, \ 0.235, \ \mathrm{and} \ 0.29 \ \mathrm{GeV}$. 
All the variables are measured in the laboratory frame of the nucleus. The data are 
from Refs.~\cite{Gothe95, Arends83}. Each plot shows the same five 
calculations as those in Fig.~\ref{fig:photopionprod_theta_60_c12}. 
Systematically with $c_{1}=1.5$, 
the $(0, \  0)$ prediction is smaller than the $(1, \  1)$ and the $(1, \  1)$
smaller than the $(1, \  0.9)$. 
In the forward kinematic region, i.e. small $\theta_{\pi}$,   
both ``$(1, \, 1), \nu=3 \ (c_{1}=3)$'' and 
``$(1, \, 0.9), \ \nu=3$'' agrees with data for the three cases. 
However for larger $\theta_{\pi}$, the calculations fail:
for $E_{\gamma}=0.235 \ \mathrm{GeV}$, the two overestimate the cross section
when $20^{\circ}\leqslant \theta_{\pi}\leqslant 60^{\circ}$ and underestimate it when $\theta_{\pi}\geqslant 60^{\circ}$; for $E_{\gamma}=0.29 \ \mathrm{GeV}$, the two give too big results compared to the data when $\theta_{\pi}\geqslant 40^{\circ}$. 
Nevertheless, we expect our calculations to work better at the higher energy region, because the cross section is more dominated by the forward production.

\subsection{Coherent pion neutrinoproduction}

\begin{figure}
\includegraphics[scale=0.6,angle=-90]{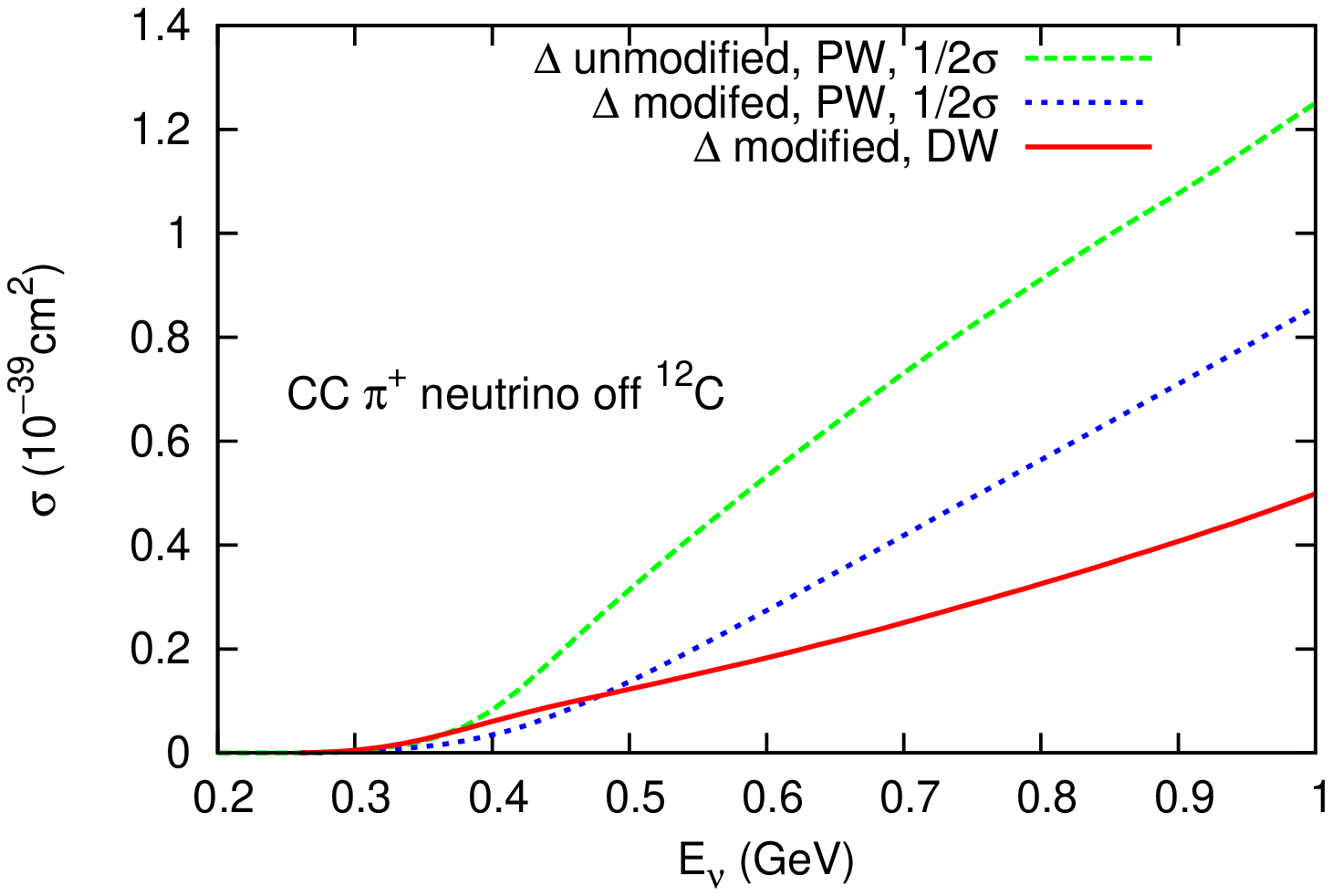}
\caption{(Color online). The total cross section for coherent CC $\pi^{+}$  production in neutrino-${}^{12}C$ scattering. 
The calculation ingredients are the same as those in Ref.~\cite{Singh06}. See the text for the explanation of different curves.}
\label{fig:checkwithPRL}
\end{figure}

Fig.~\ref{fig:checkwithPRL} shows the repeated calculations in Ref.~\cite{Singh06} for CC $\pi^{+}$ production from ${}^{12}C$. 
Only the diagrams with the $\Delta$ in $s$ and $u$ channels are included.  
We use the $N\leftrightarrow \Delta$ transition form factors in  Ref.~\cite{Singh06} to extrapolate our calculation to $E_{\nu}\geqslant 0.5$ GeV \footnote{Ref.~\cite{Singh06} labels form factors as $C_{1,2,3}^{V}$ and $C_{1,2,3,4}^{A}$.}. 
The $\Delta$ self-energy modification and the pion optical potential are also the same as in Ref.~\cite{Singh06}. This plot shows three different calculations. The ``$\Delta$ unmodified, PW, $1/2 \ \sigma$'' calculation does not apply medium-modification to the $\Delta$ self-energy; it treats the pion wave function as a plane wave; and scales the total cross section by $0.5$. In the ``$\Delta$ modified, PW, $1/2 \ \sigma$'', medium-modification for the $\Delta$ is included. Finally, the ``$\Delta$ modified, DW'' calculation includes both the medium-modification and a distorted pion wave function. A good agreement between these results and those in Ref.~\cite{Singh06} is achieved, which is a justification for our numerical calculation. 

\begin{figure}
\includegraphics[scale=0.6,angle=-90]{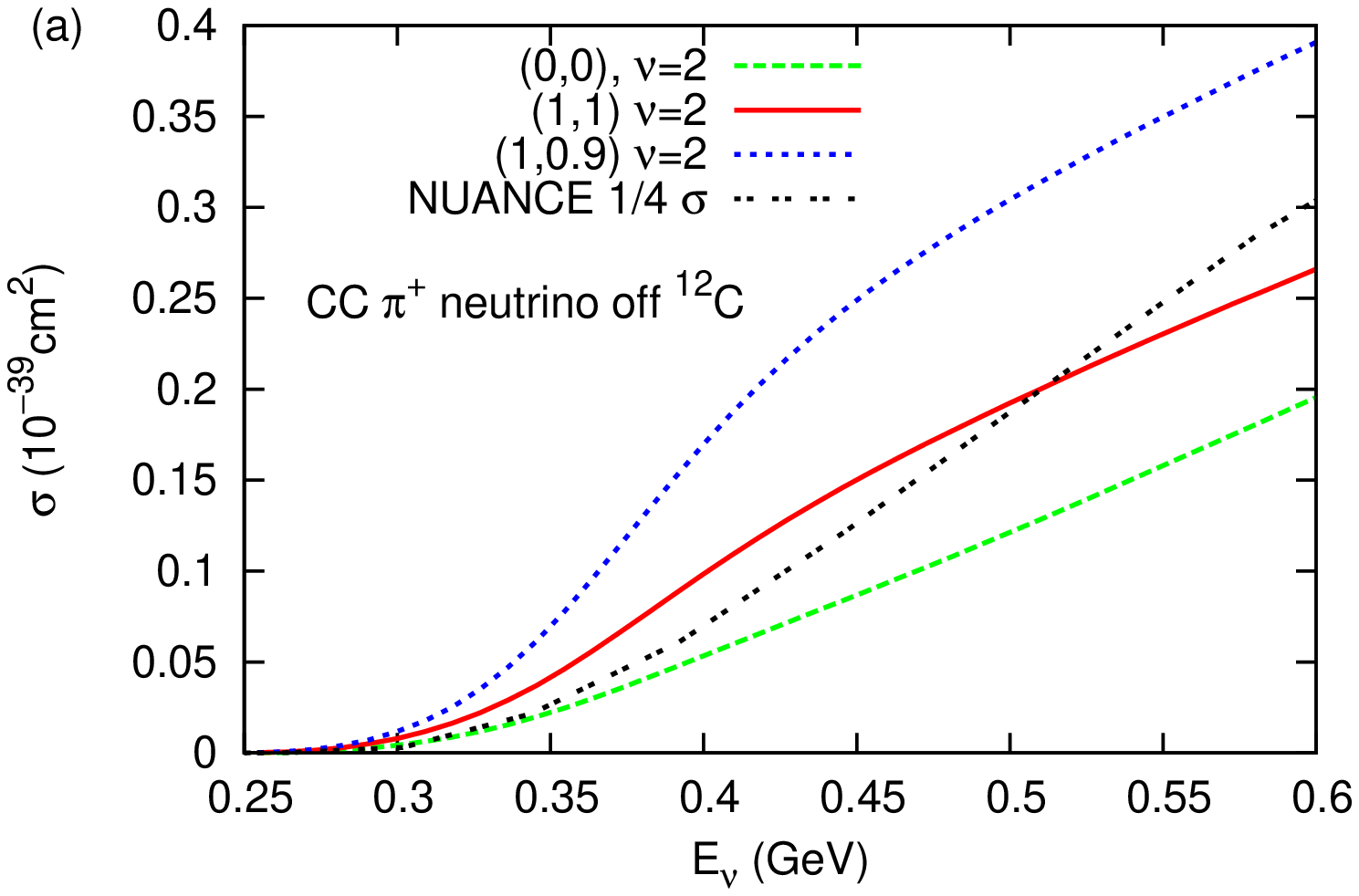}
\includegraphics[scale=0.6,angle=-90]{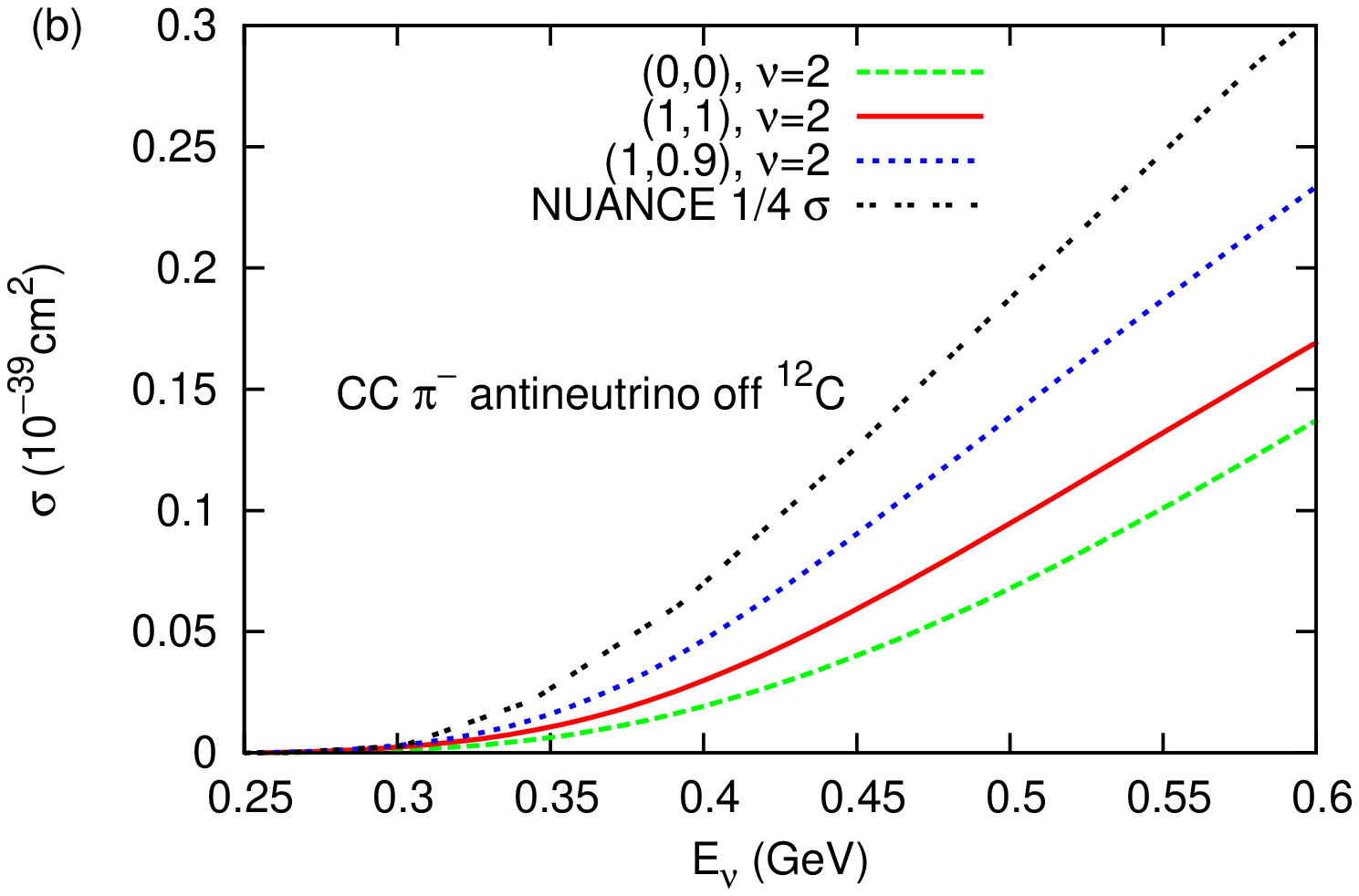}
\caption{(Color online). The total cross sections for coherent CC $\pi^{+}$ ($\pi^{-}$) production in (anti)neutrino--${}^{12}C$ scattering. 
The explanation for different curves can be found in the text.}
\label{fig:pionpm_c12_coherent}
\end{figure}

Now let's turn to our results for the total cross section of CC $\pi^{+}$ ($\pi^{-}$) production in (anti)neutrino--${}^{12}C$ scattering as shown in Fig.~\ref{fig:pionpm_c12_coherent}. Here we make use of the meson-dominance form factors that are discussed in Ref.~\cite{1stpaper} \footnote{With the meson-dominance form factors, the conservation of the  Electromagnetic currant is automatically satisfied in the free nucleon scattering calculation and in the coherent production calculation by using the optimal approximation.}. All the calculations include diagrams up to $\nu=2$ \cite{1stpaper}, with different $(r_{s}, \ r_{v})$. We also show the NUANCE output for coherent pion production, which is scaled by $1/4$. 
(In NUANCE, no wave function distortion is applied for the pion while pion absorption and rescattering are included in the subsequent step of NUANCE code \cite{Sam12}; neutrino-induced and antineutrino-induced coherent pion production have the same cross section.) In both $\pi^{+}$ and $\pi^{-}$ production, the $(1, \ 0.9)$ prediction is bigger than the $(1, \ 1)$ and the $(1, \ 1)$ bigger than the $(0, \  0)$. By comparing the differences among the three calculations with those in the incoherent productions \cite{2ndpaper}, we see the coherent processes are more sensitive to $(r_{s}$ and $r_{v})$ than the incoherent. This is consistent with the discussion in Sec.~\ref{subsec:photoprod}. Moreover, our results are much smaller than the NUANCE output in the two plots \footnote{In NUANCE, an over all $15\%$ reduction is expected after the pion absorption is included \cite{Sam12}.}. As we know, the previous calculations \cite{Rein83}, implemented by NUANCE, give bigger cross sections for coherent pion productions than the measured ones  \cite{miniboone08plb}. It is noticed in the Fig.~\ref{fig:checkwithPRL} and Refs.~\cite{Singh06,Alvarez-Ruso07CC, Alvarez-Ruso07CCErratum, Alvarez-Ruso07NC,Alvarez-Ruso07NCErratum, Amaro09, Hernandez09, Nakamura10} that including the medium-modification on the $\Delta$ and the distortion of the pion wave function reduces the cross section significantly.    

\begin{figure}
\includegraphics[scale=0.6,angle=-90]{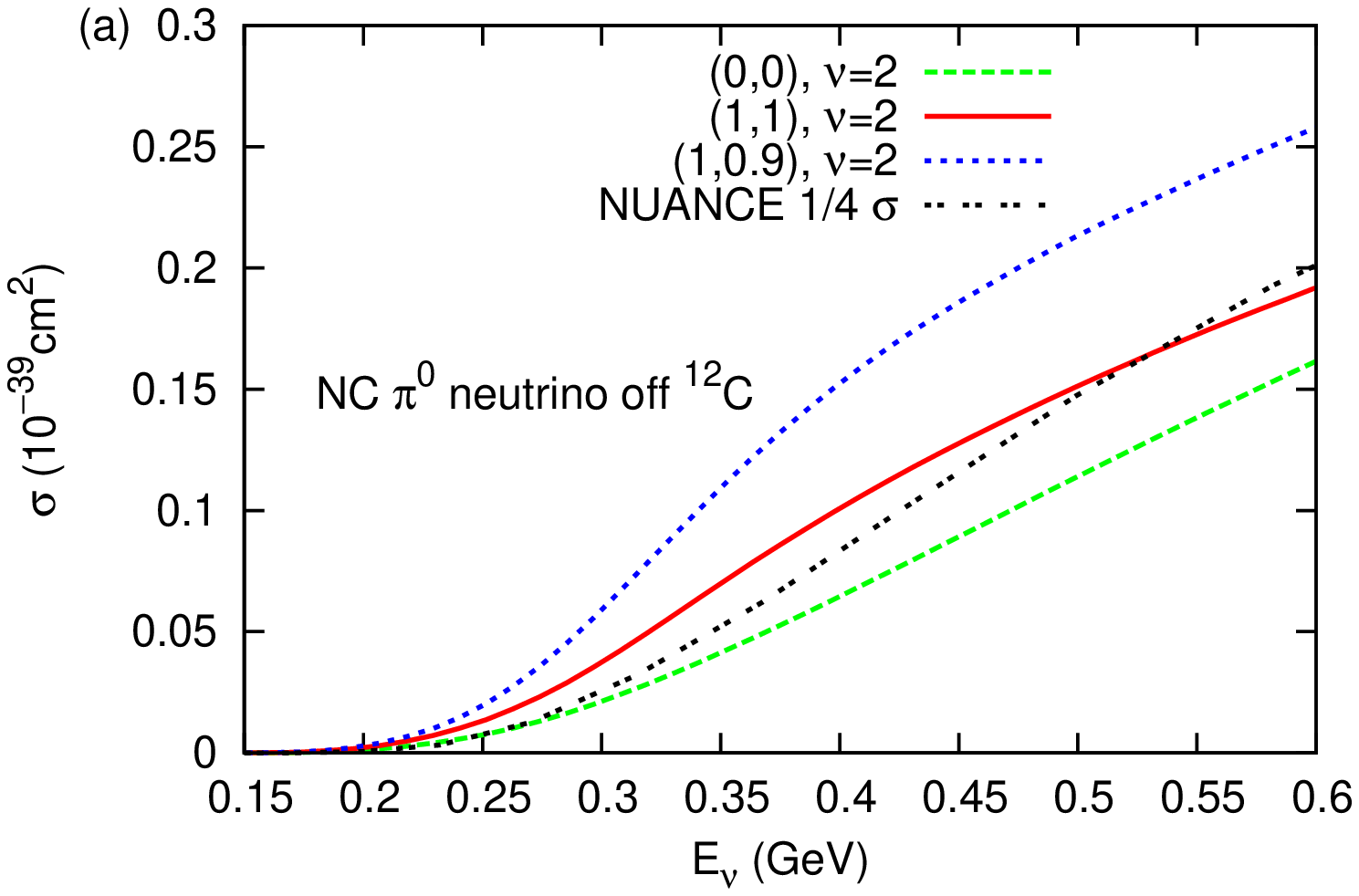}
\includegraphics[scale=0.6,angle=-90]{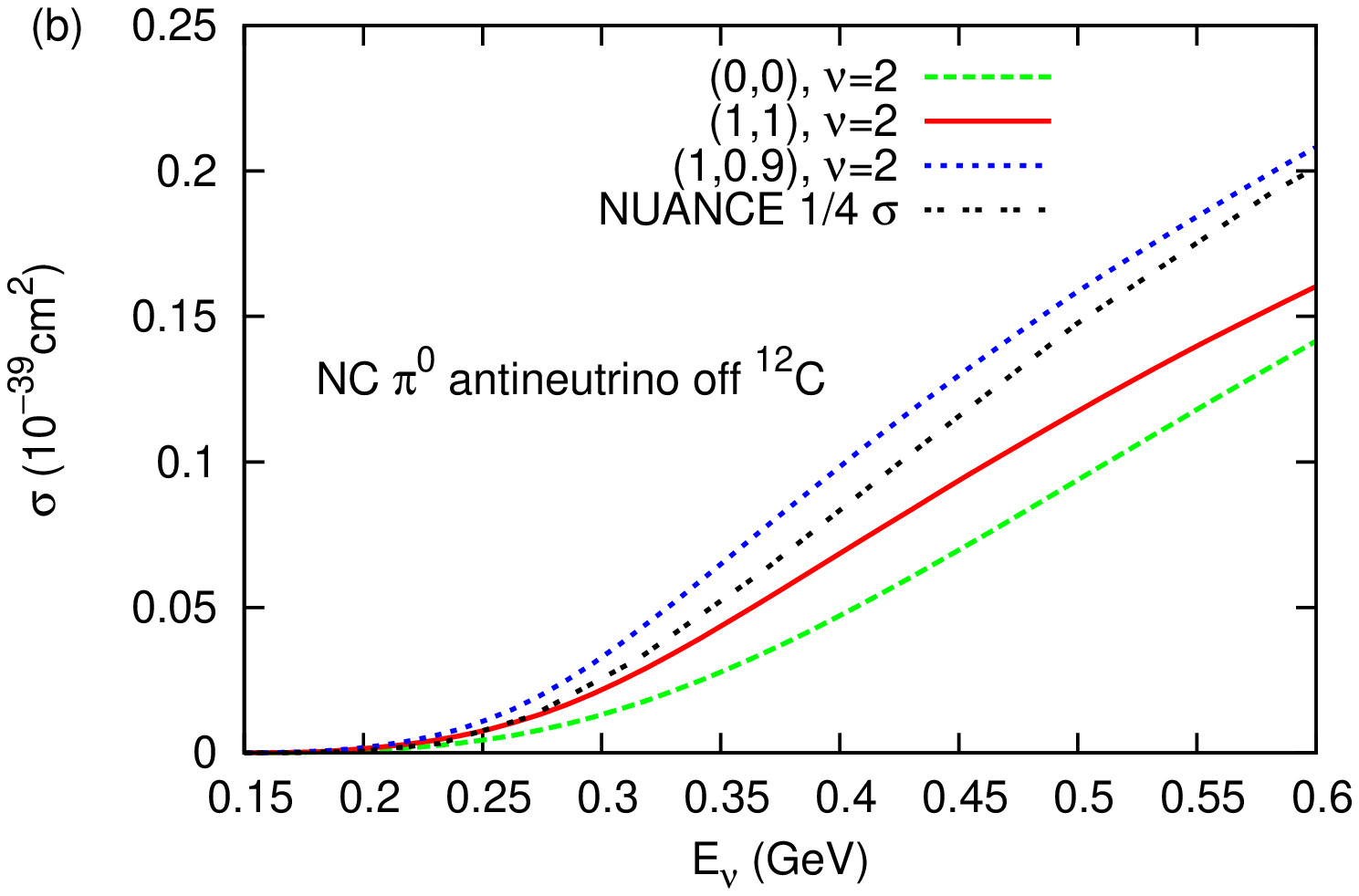}
\caption{(Color online). The total cross section for coherent NC $\pi^{0}$ production in both neutrino-- and antineutrino--${}^{12}C$ scatterings. 
The explanation for different curves can be found in the text. }
\label{fig:pion0_c12_coherent}
\end{figure}

Fig.~\ref{fig:pion0_c12_coherent} 
shows our results for NC $\pi^{0}$ production in neutrino-- and antineutrino--${}^{12}C$ scatterings. Three different calculations are presented in the same as way as in Fig.~\ref{fig:pionpm_c12_coherent}. 
The systematics in them are the same as in the CC productions: 
first, the $(1, \ 0.9)$ gives the biggest cross section and the $(0, \  0)$ gives the smallest; second, cross sections are sensitive to $r_{s}$ and $r_{v}$, compared to incoherent NC productions in Ref.~\cite{2ndpaper}; finally our results are much smaller than the NUANCE output even after including the pion absorption ($15\%$ reduction). 

In addition, as mentioned in Ref.~\cite{Rein83}, 
the coherent production is dominated by the forward production at the high energy region. By using the conservation of vector current (the leptonic current should be proportional to the momentum transfer in the forward kinematics), the contribution of the vector current in the full hadronic current is small [see Eq.~(\ref{eqn:currentpion})], and hence the interference between the vector current and axial current is small. As the result, neutrino-induced and antineutrino-induced production should have similar cross sections (see the NUANCE output shown in the plots). 
However this is clearly violated in the energy region of this paper
(see results in Fig.~\ref{fig:pionpm_c12_coherent} for the CC production 
and in Fig.~\ref{fig:pion0_c12_coherent} for the NC production). 
Furthermore, for an isoscalar nucleus like ${}^{12}C$, the axial current in the hadronic CC  and in the hadronic NC should have the same strength by using the Wigner-Eckart theorem. In the leptonic current, this ratio is $\sqrt{2} :  1$. 
Because axial current dominates in both CC and NC production at the high energy region, the ratio for the cross sections between them should be $2 :  1$ in both neutrino and antineutrino scatterings (ignore the $u$ and $d$ quark mixing) \cite{Paschos06}, which is also represented by the NUANCE output in the plots. But this ratio is not satisfied at low energy, if we compare the (anti)neutrino results in Fig.~\ref{fig:pionpm_c12_coherent} with the (anti)neutrino results in Fig.~\ref{fig:pion0_c12_coherent}. So, it will be  interesting to extrapolate our low-energy results to high energy and find out the transition region where the predicted high-energy behavior starts to emerge. 

\subsection{Coherent NC photon production}

\begin{figure}
\includegraphics[scale=0.6,angle=-90]{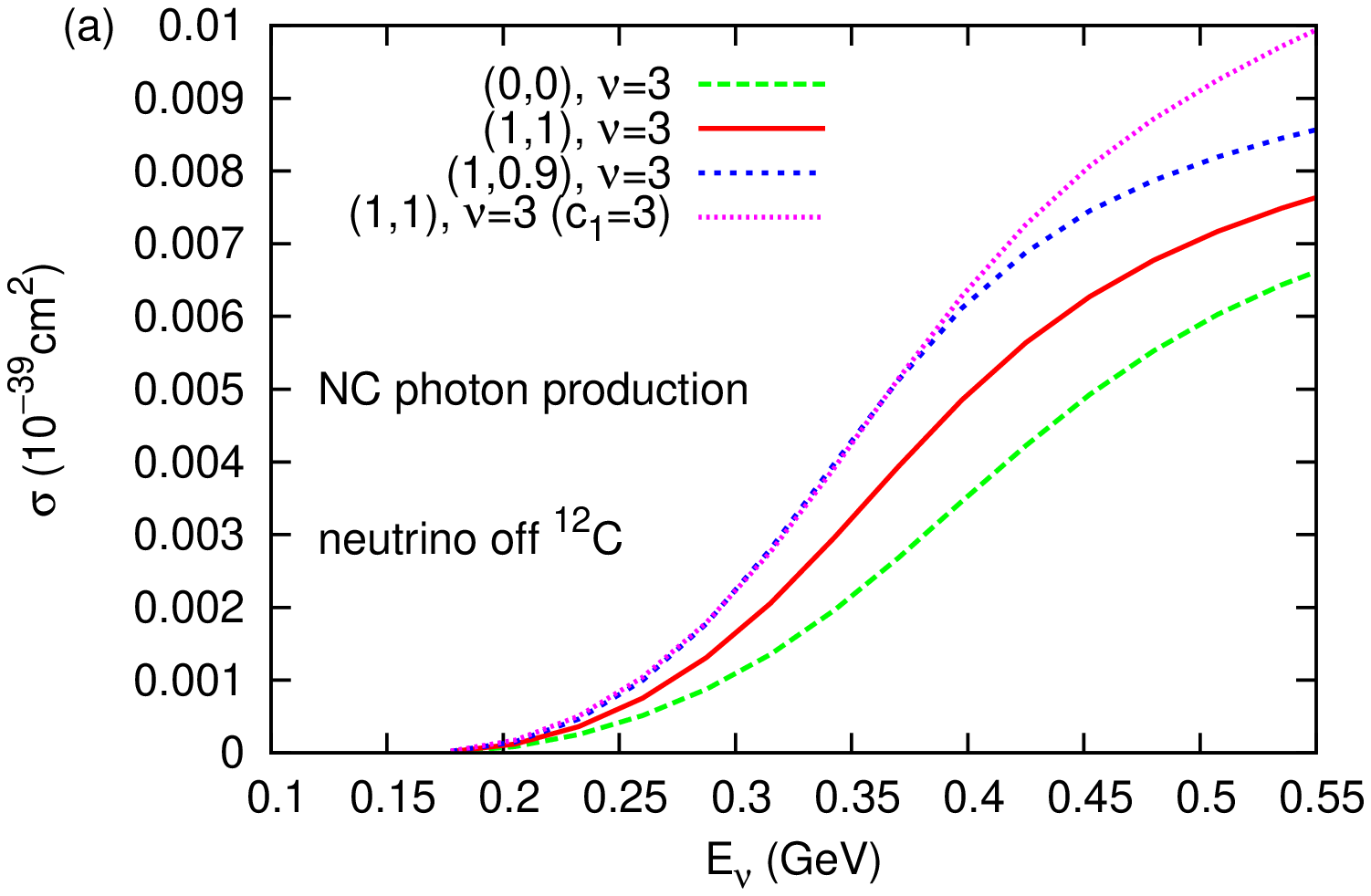}
\includegraphics[scale=0.6,angle=-90]{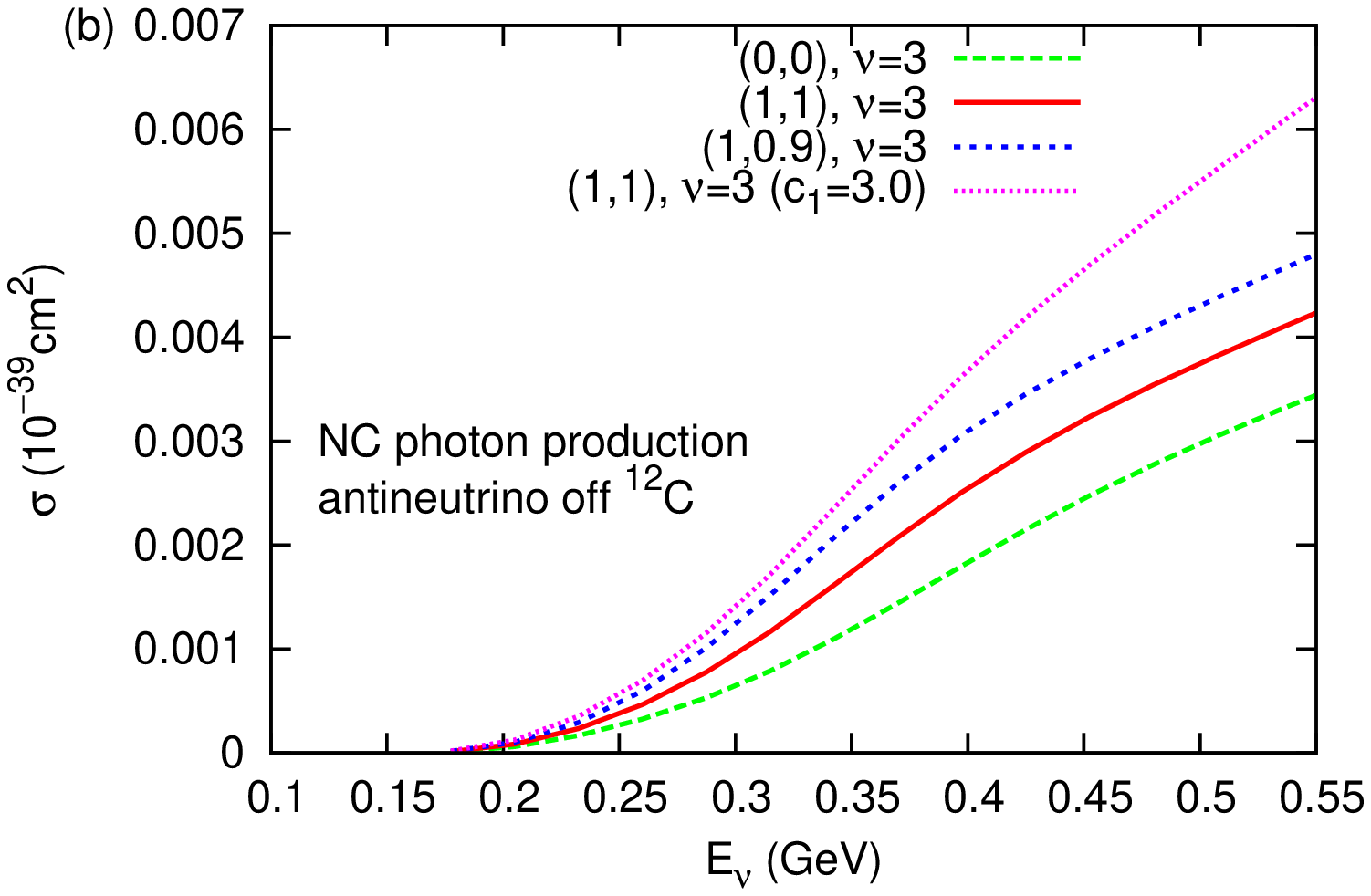}
\caption{(Color online). The total cross section for coherent NC photon production in both neutrino-- and antineutrino--${}^{12}C$ scatterings. $E_{\gamma} \geqslant 0.15 \ \mathrm{Gev}$ is applied. The explanation for different curves can be found in the text.}
\label{fig:NCphoton_c12_coherent}
\end{figure}

In Fig.~\ref{fig:NCphoton_c12_coherent}, we show our results for coherent NC  photon production in both neutrino-- and antineutrino--${}^{12}C$ scatterings \footnote{In NUANCE, there is no manifest coherent photon production channel. But in the MiniBooNE's analysis \cite{MBCCQE}, the total photon production is computed by scaling the total pion production, set from the total NC $\pi^{0}$ production data, by a proper branching ratio, which in principle has the contribution from the coherent production.}. 
In accordance with the low detection efficiency for low energy photons in 
the MiniBooNE experiment, we require the photon energy in the laboratory frame to be bigger than $0.15$ GeV in the calculation, which on the other hand simplifies the calculation because of the absence of the infrared singularity. 
The labeling of curves is the same as in pion production. 
Here all the necessary diagrams up to $\nu=3$ are included. 
As we know from Refs.~\cite{1stpaper, 2ndpaper}, all the $\nu=2$ contact diagrams do not contribute to NC photon production, and 
the $\nu=3$ diagrams are due to $c_{1}$ and $e_{1}$ coupling mentioned in the previous pion photoproduction calculation (the $e_{1}$ contribution vanishes for an isoscalar target). We can observe the effect of $c_{1}$ coupling, by comparing the ``$(1, \  1),  \nu=3$'' curve with the 
``$(1 \ , 1), \nu=3, (c_{1}=3)$'' curve [$c_{1}=1.5$ in the calculations without $(c_{1}=3)$ labeling]. It increases the total cross section by roughly $10\%$. So, at low energy the contributions of the 
contact terms are not negligible, and a similar observation is made in 
photoproduction of pions as shown in Fig.~\ref{fig:photopionprod_theta_60_c12}. \emph{But the possibility of introducing such couplings to explain the low reconstructed energy excess events in MiniBooNE is not quite promising at least considering only low-energy neutrino contributions 
\cite{Harvey07, Harvey08, Hill10}.} 
(In Ref.~\cite{MiniBN2009}, the number of excess events at low reconstructed neutrino energy is roughly two times bigger than the number of the $\Delta$ radiative decay estimated in the MiniBooNE's background analysis.)
The contributions of these terms at high energy region still need to be studied. Moreover, the hierarchy among cross sections using $(1, \  0.9)$, $(1, \ 1)$, and $(0, \  0)$ (with $c_{1}=1.5$) is also consistent with the discussion for the Fig.~\ref{fig:photopionprod_theta_60_c12}. However the 
difference among them is less significant than that in pion production, 
which is probably due to the absence of the distortion of photon wave function. The spreading between ``$(1 \ , 1), \nu=3, (c_{1}=3)$'' and ``$(1 \ , 0.9),  \nu=3$'' gives a sense of uncertainty of these calculations. 
Furthermore, the cross section in neutrino scattering is bigger than that in antineutrino scattering, which is different from the expectation about them at high energy region. 

\section{conclusion}
In this paper, we have studied coherent neutrinoproduction of photons and pions with $E_{\nu} \leqslant 0.5$ GeV. 
This paper, combined with those in Refs.~\cite{1stpaper, 2ndpaper} about the 
productions from free nucleons and the incoherent productions from nuclei, 
complete the study on these processes at low energy. 
The series is motivated by the low reconstructed energy excess events in the 
MiniBooNE experiment.   
The QHD EFT (with the $\Delta$ introduced) has been used in these works. It is a Lorentz-covariant, meson-baryon EFT with a nonlinear realization of the chiral symmetry. The $U(1)_{EM}$ gauge symmetry and chiral symmetry 
guarantee the conservation of vector current and partial conservation of axial current. These constraints seem trivial at the nucleon level, but important in many-body calculations. For example, various procedures would have to be applied by hand to make sure the vector current conserved, if gauge symmetry is not manifest. Even worse, this procedure can be entangled with the specific approximation scheme. Another advantage of working in the EFT is the power-counting of diagrams, through which we can address the relevance of some interaction vertices. In incoherent NC photon production, we see the two contact terms $c_{1}$ and $e_{1}$, which can be partially related 
with the newly proposed meson's anomalous interactions, are negligible 
(they are at next-to-next-to-leading order). Their contributions do show up 
in the coherent productions, e.g. coherent pion photoproduction and NC photon production, as demonstrated in this paper, but do not seem to increase the photon production as substantially as needed to explain the excess in the MiniBooNE experiment.  

After discussing this paper in a big context, let's proceed to summarize the specifics. The so-called optimal approximation is introduced to simplify the calculation of nuclear matrix element, in which the one-body current  matrix elements are factorized out. Meanwhile the modification on the one-body interaction amplitude is taken into account. The real part of the nucleon and the $\Delta$ self-energies is calculated by using the mean-field approximation of this model. The change of the $\Delta$ width is parameterized in a phenomenological way according to pion-nucleus scattering data. The medium-modifications have been tested in incoherent pion production in Ref.~\cite{2ndpaper}. The Eikonal approximation is used to handle the distortion of the final pion wave function. Moreover, we have compared our 
approximation with the one used in \cite{AbuRaddad99} in which the authors
introduce other densities besides the baryon density used in our approximation. It is shown in Fig.~\ref{fig:compared_with_currentdecomposed} that the two methods give similar results. 

We calculate the differential cross sections for pion photoproduction, which serves as the benchmark for our approximations, and then calculate the total cross sections for pion neutrinoproductions. The results are sensitive to $r_{s}$, $r_{v}$, and the contact term $c_{1}$. The disagreement at high energy with the fixed pion angle shown in Fig.~\ref{fig:photopionprod_theta_60_c12} and at the big pion angle with fixed photon energy shown in Fig.~\ref{fig:photopionprod_Ei_c12} seems to indicate that it is necessary to go beyond the one-body current approximation to explain the full data. However, to resolve the disagreement, both the $\Delta$ dynamics and the  distortion of the pion wave function should be understood better as well. In addition, we also compare our neutrinoproduction results with those in the literature to check our numerical calculation. Finally in photon neutrinoproduction, the total cross sections also depend on 
$r_{s}$, $r_{v}$, and $c_{1}$. Changing $c_{1}$ from 1.5 to 3 increases both  neutrino- and antineutrino-induced photon production by roughly $10\%$. 
  
Now, let's come back to the question about the photon production being the excess events 
in the MiniBooNE experiment. One tricky point should be pointed out here. The 
\emph{reconstructed} neutrino energy is based on CC quasi elastic scattering 
kinematics, which can underestimate the neutrino energy in the photon production.  
So the high energy neutrino contribution to the photon production should be addressed before drawing a definite conclusion for this question. 
The calculations in Ref.~\cite{2ndpaper} and this paper illustrate the approximations used in both incoherent and coherent productions, and provide important calibration for the modification of the one-nucleon interaction amplitude in nuclei. The sensible extrapolations of current results to the high energy region will be pursued in future work.  

\acknowledgments
XZ would like to thank Joe Grange, Teppei Katori, William C. Louis, Rex  Tayloe, and Geralyn Zeller for their valuable information and useful discussions. This work was supported by the Department of Energy under Contract No.\ DE--FG02--87ER40365.

\end{document}